\documentclass[useAMS,usenatbib]{mn2e}

\usepackage{times,graphicx,amssymb,natbib}

\title[Stellar Encounters in the Context of Outburst Phenomena]{Stellar Encounters in the Context of Outburst Phenomena}
\author[Duncan Forgan and Ken Rice]{Duncan Forgan $^{1}$\thanks{E-mail:
dhf@roe.ac.uk} and Ken Rice$^{1}$\\
$^{1}$Scottish Universities Physics Alliance (SUPA), Institute for Astronomy, University of Edinburgh, Blackford Hill, Edinburgh, EH9 3HJ, Scotland, UK \\
}
\begin{document}

\date{Accepted 0000}

\pagerange{\pageref{firstpage}--\pageref{lastpage}} \pubyear{0000}

\maketitle

\label{firstpage}

\begin{abstract}

\noindent Young stellar systems are known to undergo outbursts, where the star experiences an increased accretion rate, and the system's luminosity increases accordingly.  The archetype is the FU Orionis (FU Ori) outburst, where the accretion rate can increase by three orders of magnitude (and the brightness of the system by five magnitudes).  The cause appears to be instability in the circumstellar disc, but there is currently some debate as to the nature of this instability (e.g. thermal, gravitational, magneto-rotational).

This paper details high resolution Smoothed Particle Hydrodynamics (SPH) simulations that were carried out to investigate the influence of stellar encounters on disc dynamics.  Star-star encounters (where the primary has a self-gravitating, marginally stable protostellar disc) were simulated with various orbital parameters to investigate the resulting disc structure and dynamics.  Crucially, the simulations include the effects of radiative transfer to realistically model the resulting thermodynamics.

Our results show that the accretion history and luminosity of the system during the encounter displays many of the features of outburst phenomena.  In particular, the magnitudes and decay times seen are comparable to those of FU Ori.  There are two caveats to this assertion: the first is that these events are not expected to occur frequently enough to explain all FU Ori or EX Lupi; the second is that the inner discs of these simulations are subject to numerical viscosity, which will act to reduce the accretion rate (although it has less of an effect on the total mass accreted).  In short, these results cannot rule out binary interactions as a potential source of some FU Ori-esque outbursts.

\end{abstract}


\section{Introduction}

The traditional picture of star formation describes the free-fall collapse of a protostellar molecular cloud (e.g. \citealt{Shu_et_al_87}).  Conservation of angular momentum will in general ensure that, within typical free-fall times of \(\sim 10^5\) yr, the collapse produces a protostar with protostellar disc.  These formation rates are consistent with the observed statistics of protostellar objects, for example those in Taurus \citep{Kenyon_et_al_90} and with numerical simulations \citep{Bate_cluster_03,Stam_frag,Bate_09}.  If these formation timescales are converted to average mass accretion rates, then it appears that the standard picture will form a star at the rate of \(10^{-5} M_{\odot}\, yr^{-1}\), whereas current observations suggests an average infall rate of \(10^{-6} M_{\odot}\, yr^{-1}\) or lower.  This leads to the realisation that accretion rates in protostars are not constant, which has been confirmed observationally \citep{Herbig_77, Armitage_et_al_01,Zhu_et_al_09}, and that short periods of increased accretion, (accompanied by periods of mass pile-up where the infalling matter is not accreted) can solve the apparent inconsistencies between observation and theory.

This also provides an explanation for the FU Orionis (FU Ori) outburst objects, Class 0 to Class II protostellar objects which undergo a characteristic rapid rise in luminosity spanning up to five magnitudes, and then decay on timescales of a few hundred years \citep{Hartmann_Kenyon_96}.  Calculated accretion rates for FU Ori objects are typically \(10^{-4} M_{\odot}\, yr^{-1}\) or more \citep{Herbig_77,Hartmann_Kenyon_96}, which show a strong increase over typical infall rates \citep{Kenyon_et_al_93,Furlan_et_al_08}.

The precise origin of these FU Ori outbursts is not known, although almost all theories involve a protostellar disc, as these discs are expected to be present around the majority of early-type stars.  The theories range from thermal instability \citep{Lin_Papa_85,Bell_and_Lin}, gravitational instability \citep{Vorobyov_Basu_05,Vorobyov_Basu_06,Vorobyov_Basu_08,Boley_and_Durisen_09}, a combination of gravitational instability and magnetorotational instability \citep{Armitage_et_al_01,Zhu_et_al_09}, even the presence of a planet controlling accretion onto the central star \citep{Clarke_Syer_96,Lodato_Clarke_04}.  All these theories agree that triggering a disc instability event is crucial to providing the observed accretion rates that cause the outburst.  The aim of this paper is not to identify the correct theory of FU Ori objects, but to comment on what was once considered a potential cause of FU Ori: the encounter of a primary star plus protostellar disc with a discless secondary \citep{Kenyon_et_al_88,Bonnell_Bastien_92}.  

Observations are beginning to reveal that outburst phenomena appear to belong to different classes, e.g. FU Ori (FUors), EX Lupi (EXors) \citep{Herbig_07}, and others.  This work proposes that stellar encounters could produce outbursts that share many observational features with FUors and EXors (as was suggested by \citet{Pfalzner_08}, combining treecode simulations with cluster dynamics), and yet may belong to a different class.   Results are presented from a series of high resolution Smoothed Particle Hydrodynamics (SPH) simulations with radiative transfer \citep{intro_hybrid} of the encounter of a star-disc system with a discless companion.  The luminosity of these systems during the encounter is studied, and their potential for observation as a separate subclass of eruptive variable is discussed.

\section{Method} \label{sec:method}

\begin{table*}
\centering
  \caption{Summary of the orbital parameters investigated in this work.\label{tab:params}}
  \begin{tabular}{c || cccccccc}
  \hline
  \hline
   Simulation  &  $M_{disc}/M_{\odot}$   & $\Sigma \propto r^{-x}$  &  $M_{2}/M_{\odot}$   & Calculated $R_{peri}$ (au)  & Actual $R_{peri}$ (au)& $e$ &   Prograde/Retrograde &   Inclination    \\  
 \hline
  1 & 0.1 & 1 & 0.1 & 40 & 28 & 1 & Pro & $0^{\circ}$ \\
 2 & 0.1 & 1 & 0.1 & 30 & 25 & 1 & Pro & $0^{\circ}$ \\
  3 & 0.2 & 1 & 0.1 & 50 & 36 &1 & Pro & $0^{\circ}$ \\
  7 & 0.1 & 1 & 0.1 & 30 & 33 & 7 & Pro & $0^{\circ}$ \\
 \hline
  \hline
\end{tabular}
\end{table*}

\subsection{SPH and the Hybrid Radiative Transfer Approximation}

\noindent Smoothed Particle Hydrodynamics (SPH) \citep{Lucy,Gingold_Monaghan,Monaghan_92} is a Lagrangian formalism that represents a fluid by a distribution of particles.  Each particle is assigned a mass, position, internal energy and velocity: state variables such as density and pressure can then be calculated by interpolation - see reviews by \citet{Monaghan_92,Monaghan_05}.  In these simulations, the gas is modelled by 500,000 SPH particles: the primary star (and the secondary companion) are represented by point mass particles, which can accrete gas particles if they are sufficiently close and are bound \citep{Bate_code}.

The SPH code used in this work is based on the SPH code developed by \citet{Bate_code}.  It uses individual particle timesteps, and individually variable smoothing lengths \(h_i\) such that the number of nearest neighbours for each particle is \(50 \pm 20\).  The code uses a hybrid method of approximate radiative transfer \citep{intro_hybrid}, which is built on two pre-existing radiative algorithms: the polytropic cooling approximation devised by \citet{Stam_2007}, and flux-limited diffusion (e.g. \citealt{WB_1,Mayer_et_al_07}, see \citealt{intro_hybrid} for details).  This union allows the effects of both global cooling and radiative transport to be modelled without extra boundary conditions. 

The opacity and temperature of the gas is calculated using a non-trivial equation of state - this accounts for the effects of H$_{2}$ dissociation, H$^{0}$ ionisation, He$^{0}$ and He$^{+}$ ionisation, ice evaporation, dust sublimation, molecular absorption, bound-free and free-free transitions and electron scattering \citep{Bell_and_Lin,Boley_hydrogen,Stam_2007}.  Heating of the disc can also be achieved by \(P\,dV\) work, shocks and viscous dissipation.

\subsection{Initial Disc Conditions}

\noindent The discs used in this work were initially evolved in isolation for several Outer Rotation Periods (ORPs).  This allows the disc to approach an equilibrium state and become marginally unstable, and to develop steady-state spiral structures \citep{Lodato_and_Rice_04}.  The discs extend from \(r_{in} = 1\) au to \(r_{out} = 40 \) au.  The disc used for most of the simulations described here has a mass of \(0.1 \,M_{\odot}\) with a central primary star of mass \(0.5 \,M_{\odot}\).  The initial surface density profile was chosen to be \(\Sigma \propto r^{-1}\), with a sound speed profile of \(c_s \propto r^{-\frac{1}{4}}\).  One variant of the disc was also run with a disc mass of \(0.2 \, M_{\odot}\).  These initial conditions (in particular the small disc radii) were motivated by recent work on disc fragmentation, which suggests that massive discs with radii of order $\sim$ 100 au and greater will tend to fragment in the outer regions, on timescales close to the dynamical timescale \citep{Stam_frag,Clarke_09,Rice_and_Armitage_09}.  This is consistent with observations that massive discs tend to have outer radii less than 100 au \citep{Rodriguez_et_al_05}.

\begin{figure*}
\begin{center}$
\begin{array}{cc}
\includegraphics[scale = 0.5]{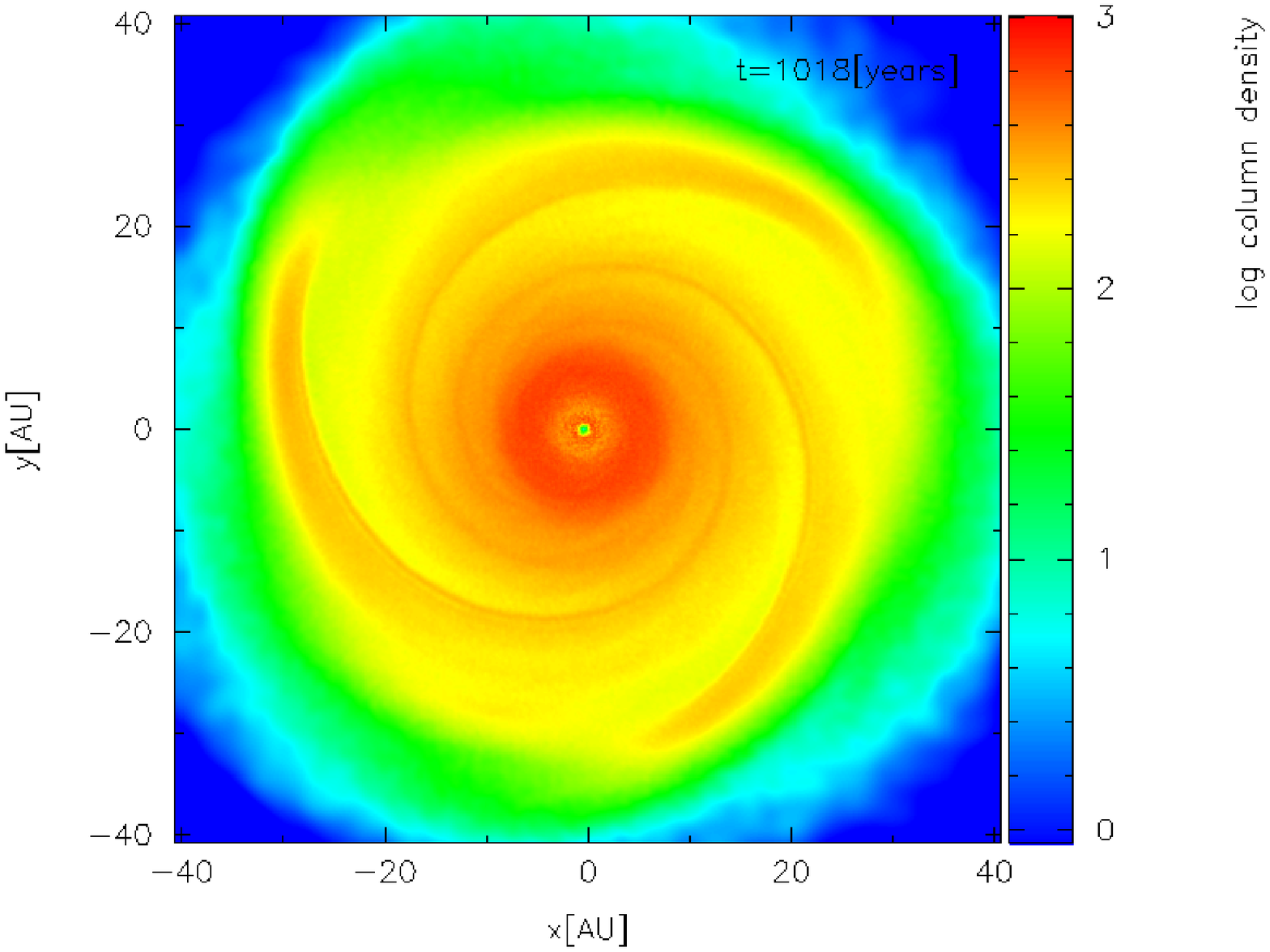} \\
\includegraphics[scale = 0.5]{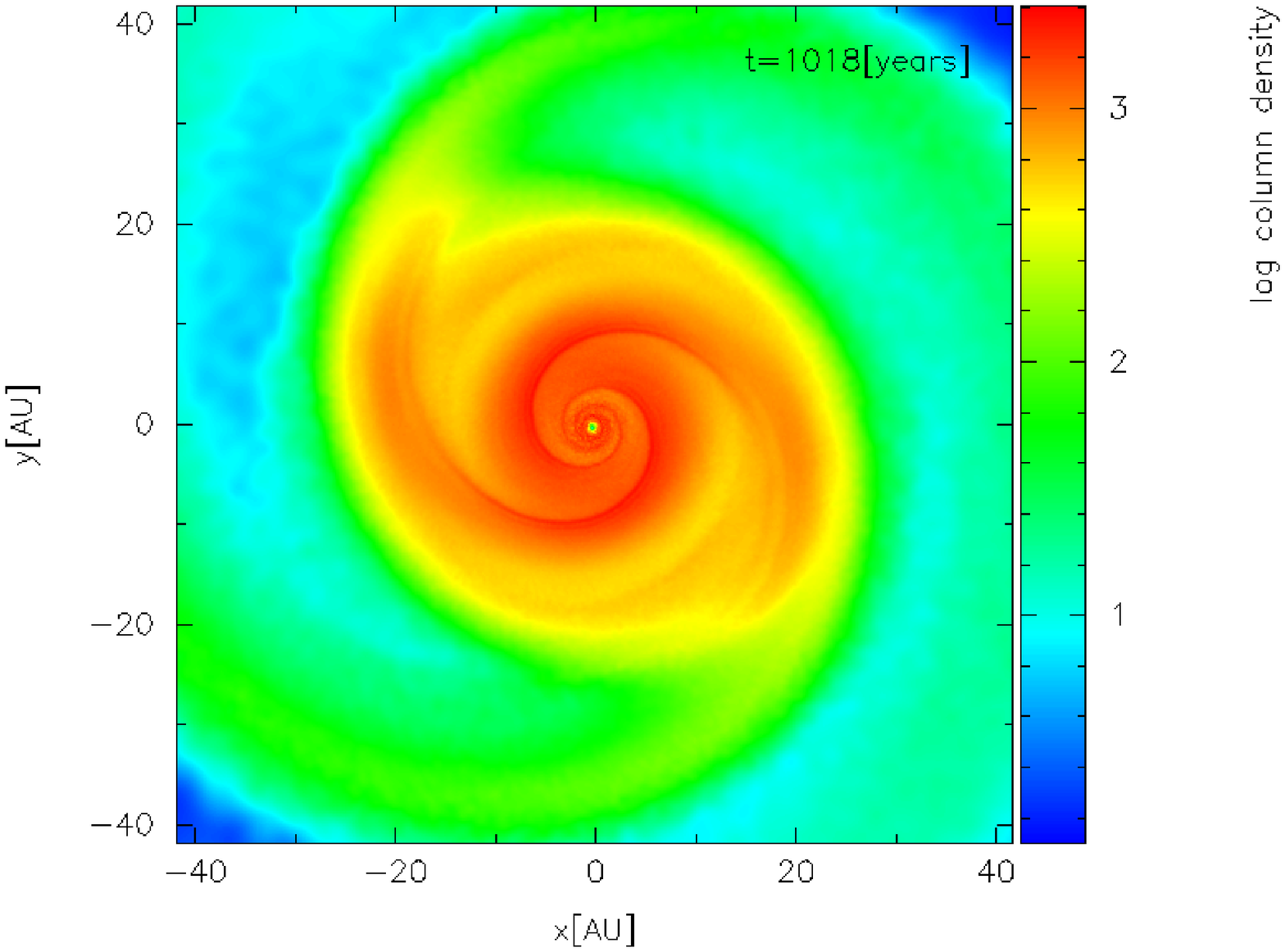} \\
\end{array}$
\caption{Snapshots of the two discs used after several ORPs: the \(0.1\, M_{\odot}\) disc (top), and the \(0.2\, M_{\odot}\) disc (bottom).\label{fig:disc}}
\end{center}
\end{figure*}

\begin{figure*}
\begin{center}$
\begin{array}{cc}
\includegraphics[scale = 0.5]{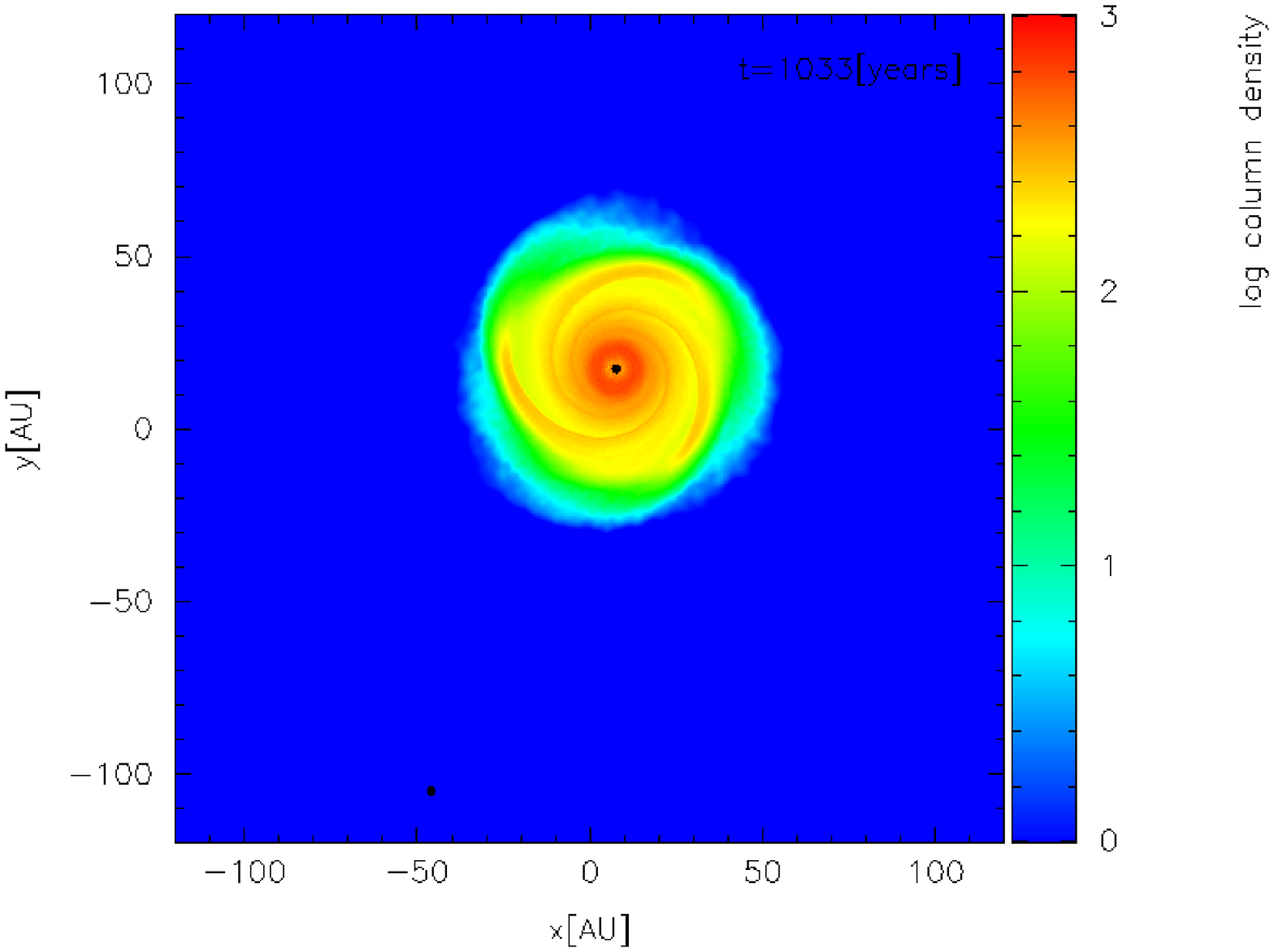} \\
\includegraphics[scale = 0.5]{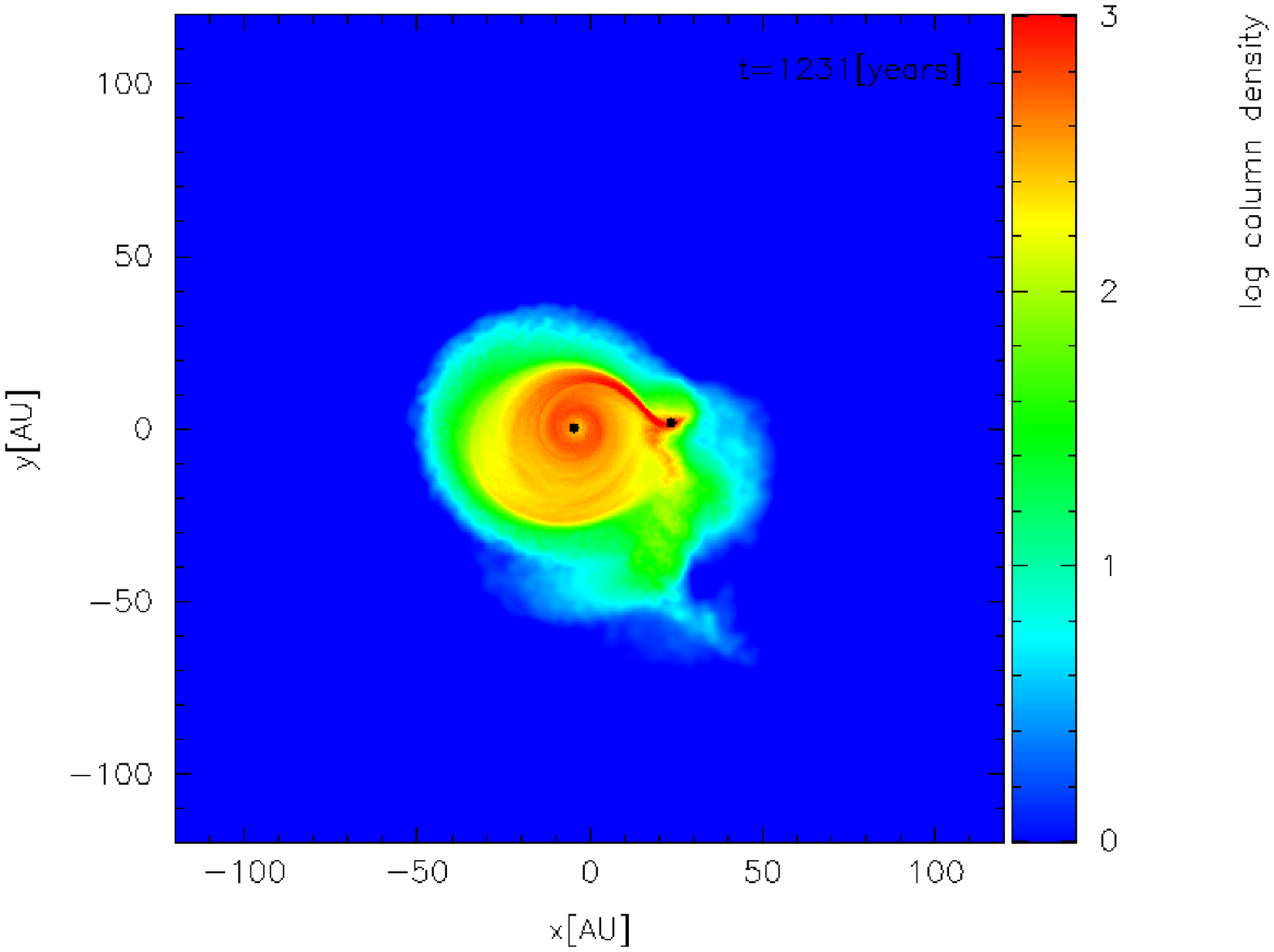} \\
\includegraphics[scale = 0.5]{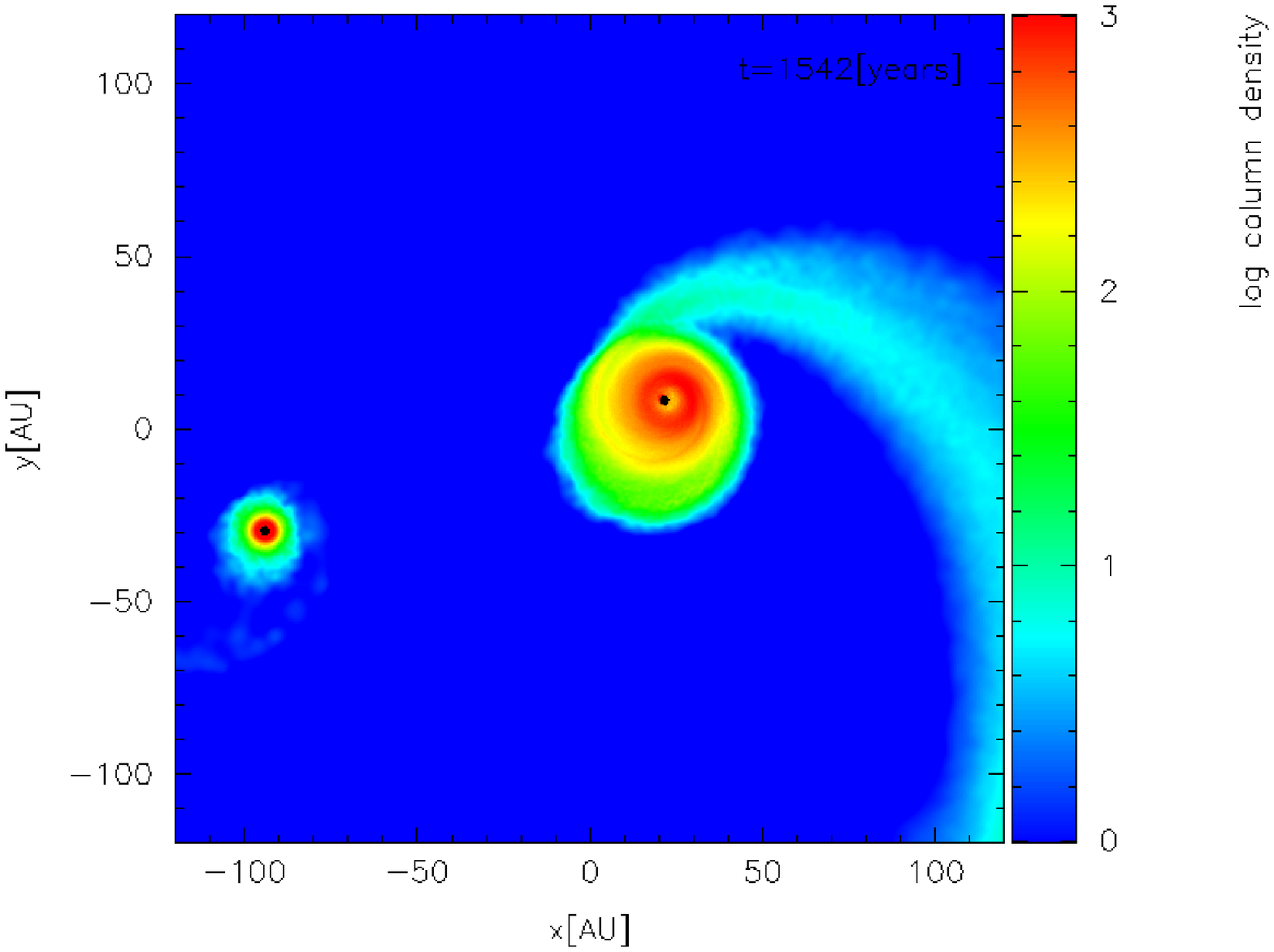} \\
\end{array}$
\caption{Images of the Simulation 1 disc before, during, and after the encounter (dot-dashed line). \label{fig:image_1}}
\end{center}
\end{figure*}	 

\subsection{The Stimulus: Adding a Companion}

\noindent With the disc evolved into a quasi-steady state, a secondary star was then added (at a separation of at least 100 au to the primary, to prevent any non-linear perturbations in the disc by the secondary's sudden appearance).  The secondary was added with several different initial conditions, such that the system exhibited different semimajor axes and eccentricities, comprising a suite of 4 simulations (see Table \ref{tab:params} for details).  All encounters are prograde, and coplanar to the disc: all encounters are also parabolic (with the exception of simulation 4, which has an eccentricity of 7).  This allows the characterisation of how outburst behaviour is dependent on initial orbital parameters. 
 
\subsection{Resolution}

\noindent It is of utmost importance that SPH simulations involving mass accretion appropriately resolve mass elements.  To correctly model the accretion of matter onto both masses, the available feedstock for accretion must be identified.  The primary can accrete from the inner regions of the disc, which are the densest (and hence best represented by SPH particles).  In the innermost annulus (say up to 0.5 au from the disc's inner edge), there is 0.015 \(M_{\odot}\) of material (given a disc of mass 0.1 \(M_{\odot}\) and a surface density profile \(\Sigma \propto r^{-1}\)).  These simulations use 500,000 SPH particles: this mass therefore corresponds to around 80,000 particles.  The minimum mass element that SPH can resolve is typically one nearest neighbour group (e.g. \citealt{Burkert_Jeans}), which in this work corresponds to 50 particles. Therefore, the accretion feedstock for the primary constitutes around 1600 nearest neighbour groups, comfortably above the minimum mass resolution.  The problem of increasing numerical viscosity in the inner regions is unfortunately insurmountable \citep{Clarke_09}.  The artifically high viscosity in the inner regions will prevent mass piling up (e.g. \citealt{Rice_and_Armitage_09}), so it should be expected that the primary's accretion rate will be underestimated (although the total mass accreted may not be affected, as the accretion will begin earlier, and for a longer duration).  Comparing the contributions to the viscosity parameter $\alpha$ \citep{Shakura_Sunyaev_73} from Reynolds stresses and from gravity shows that artificial viscosity dominates these discs within the inner 10 au. At $R=10$ au, $\alpha_{Reyn} = \alpha_{grav} \sim 10^{-3}$. 

The secondary can accrete from matter it encounters along its trajectory: the matter must come sufficiently close to become bound to the secondary before accretion is possible.  This defines the secondary's feedstock locale (approximately) as a semi-annulus in the disc, centred on periastron, with upper and lower radial boundaries based on the secondary's gravitational influence.  It is a semi-annulus because only disc material in the correct orbital phase will come into close proximity with the secondary.  Material on the opposite side of the disc during the encounter is typically teased into a tidal tail that is not accreted by either body.

In more rigorous terms, a gas element \(i\) must satisfy the following for capture by the secondary:

\begin{equation} E = E_{kin} + E_{pot} = \frac{1}{2} m_i v_{i2}^2 - \frac{GM_2 m_i}{r_{i2}} < 0, \label{eq:captureE}\end{equation}

\noindent where \(M_2\) is the secondary mass, \(r_{i2}\) and \(v_{i2}\) are the position and velocity of the gas element relative to the secondary.  Assuming that the majority of capture occurs around periastron, then 

\begin{equation} v_{i2} \approx |v_{i1} - v_{2,peri}| \end{equation}
\begin{equation} r_{i2} \approx |r_{i1} - r_{2,peri}| \end{equation}

\noindent where \(r_{i1}\) is the separation between the gas element and the primary, and \(v_{i1}\) is the velocity of the gas relative to the primary.  If the disc is Keplerian (and no radial motion is assumed), then

\begin{equation} v_i1 = r_{i1} \Omega_{i1} = \sqrt{\frac{GM_1}{r_{i1}}}, \end{equation}

\noindent The standard orbital equations can be used for \(v_{2,peri}\).  This defines the semi-annulus in the primary disc, where the secondary exerts sufficient influence to potentially capture disc material.  The upper and lower limits of this semi-annulus are found numerically by solving

\begin{equation} \frac{1}{2} \left(\sqrt{\frac{GM_1}{r_{i1}}} - \sqrt{\frac{2GM_1}{r_{2,peri}}}\right)^2 - \frac{GM_2}{r_{i2}} = 0 \label{eq:rbound}\end{equation}

Having defined the semi-annulus, it is a simple matter to calculate its mass, given a surface density profile for the disc.  In the case of a \(\Sigma \propto r^{-1}\) disc with mass 0.1 \(M_{\odot}\), this gives an available mass of \(\sim\) 0.0057 \(M_{\odot}\).  Again, the simulations use 500,000 SPH particles, and hence this corresponds to a particle number of \(\sim\) 30,000.  Therefore, the accretion feedstock for the secondary constitutes 600 nearest neighbour groups, again well above the minimum mass resolution. 

\section{Results}\label{sec:results}

\subsection{Simulation 1 - The Reference Simulation}

\noindent Images of the reference simulation can be seen in Figure \ref{fig:image_1}.  As the secondary passes through the disc, it imparts significant energy to the disc \citep{Lodato_encounters,encounters}.  It captures a significant secondary disc (roughly 0.006 \(M_{\odot}\)) during the encounter (slightly more than expected from the analysis given in the previous section).  This is due to the tidal forces exerted by the disc on the secondary as it reaches periastron, reducing its velocity and allowing more matter to be captured.  The encounter draws out a significant tidal tail, and erases the spiral structure previously seen.

\begin{figure}
\begin{center}
\includegraphics[scale = 0.5]{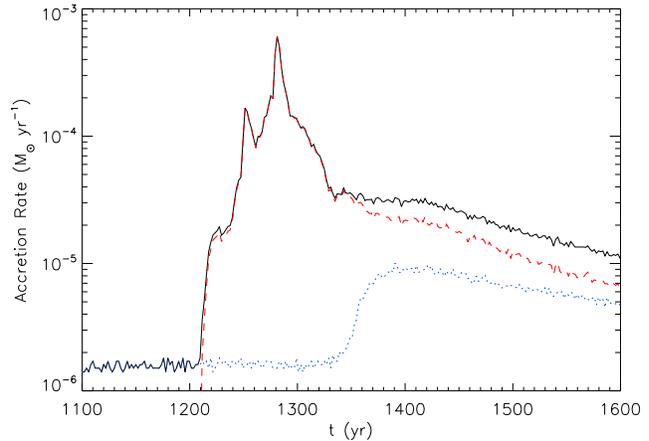} \\
\caption{Accretion rates for the primary and secondary in Simulation 1.  The black line denotes total mass accretion, the blue line indicates the primary accretion, the red line indicates secondary accretion.  \label{fig:mdot_1}}
\end{center}
\end{figure}

\begin{figure}
\begin{center}
\includegraphics[scale = 0.5]{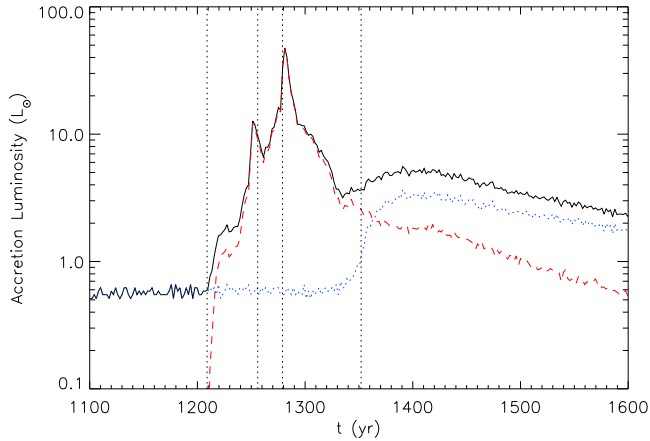} \\
\caption{Accretion luminosities for the primary and secondary (and disc luminosity) in Simulation 1.  The black line denotes total luminosity, the blue line indicates the primary accretion, the red line indicates secondary accretion, and the green line indicates the disc luminosity.  \label{fig:ptlum_1}}
\end{center}
\end{figure}

The accretion rate of the primary and secondary can be seen in Figure \ref{fig:mdot_1}, with corresponding accretion luminosities in Figure \ref{fig:ptlum_1}.  The accretion luminosity is calculated using

\begin{equation} L_{acc} = \frac{1}{2} \frac{GM\dot{M}}{R} \end{equation}

\noindent where R indicates the accretion radius of the object (taken for both objects to be 0.1 AU in these simulations). Particles are accreted if they pass within the accretion radius of the object, and are gravitationally bound.  R is held constant throughout the simulation, only \(M\) and \(\dot{M}\) change.  As the luminosity varies linearly with the accretion rate, the accretion luminosities show similar peaks and troughs as the accretion rates (with the important modification that the primary's larger mass makes it more luminous).

As a guide, Figure \ref{fig:ptlum_1} has several important events delineated by vertical lines:

\begin{itemize}
\item t \(\sim\) 1209 yr - the secondary begins its ingress to the disc, establishing a non-zero accretion rate.
\item t \(\sim\) 1256 yr - the secondary reaches periastron.  The disc's temperature also peaks at this time.
\item t \(\sim\) 1279 yr - The secondary reaches its peak accretion rate: the matter that forms the secondary disc (located in the outer tidal tail generated at the location of the secondary) is in the process of infall onto its new parent star, not taking a well defined spheroidal shape until it has begun to exit the disc.  The disc's luminosity has returned to near pre-encounter levels.
\item t \(\sim\) 1352 yr - the secondary begins its egress from the disc, and its gravitational influence diminishes.  The primary disc must now begin readjustment - an increase in outward angular momentum transport gives a corresponding increase in inward mass flux, boosting the primary's accretion rate.  The rate has a rise time of approximately 25 years.
\item t \(>\) 1352 yr - the luminosities decay with timescales of hundreds of years.
\end{itemize}

\noindent It must be emphasised that there are two separate accretion events, with distinct characteristics: the secondary's accretion undergoes enhancement for a period of around 150 years, reaching a maximum accretion rate of \(5 \times 10^{-4} \, M_{\odot} yr^{-1}\) at periastron.  Its accretion rate curve is a superposition of three characteristic features:

\begin{enumerate}
\item A smooth symmetric curve, width approximately 110 years.  This is linked to the secondary's passage through the smooth component of the disc.
\item A series of small spikes, widths of 1 year or less.  These are due to the secondary's passage through overdensities caused by the disc's spiral structure.
\item A major spike, width approximately 10 years.  This is caused by the formation of the secondary disc.
\end{enumerate}

The primary's accretion curve consists of a steady increase to values of \(\sim 10^{-5} \, M_{\odot} yr^{-1}\) as the secondary leaves the disc, with a rise time of order $\sim$ 10 years.  This is followed by a slow decay with timescales in excess of 300 years.  The accretion curves of the subsequent simulations discussed share some or all of these characteristic features.

It should again be noted that these simulations are subject to numerical viscosity in their inner regions, preventing mass build up. Some theories suggest that this pile up is crucial to activating MRI, which causes the outburst \citep{Armitage_et_al_01,Zhu_et_al_09}.  The artificially high viscosity in the inner regions can be thought of as acting like MRI, which activates at the ``wrong'' temperatures, facilitating mass accretion without pile up, and hence underestimating accretion rates during the outburst.  These simulations can then be thought of as presenting the lower bounds of the accretion rate (in the limit where MRI is overactive). 

\begin{figure}
\begin{center}
\includegraphics[scale = 0.5]{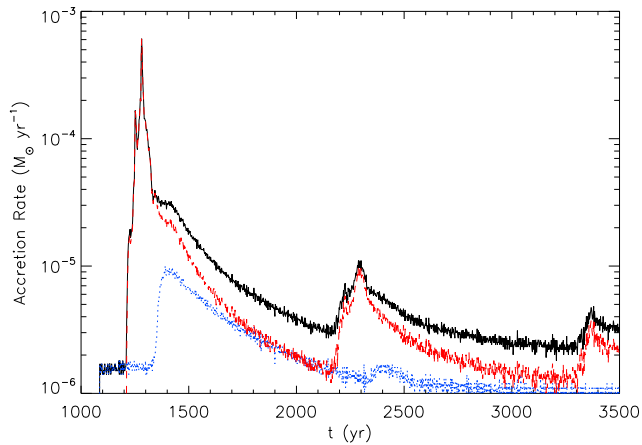} \\
\caption{Long term evolution of the accretion rates for the primary and secondary in Simulation 1.  The black line denotes total mass accretion, the blue line indicates the primary accretion, the red line indicates secondary accretion.  \label{fig:mdot_1long}}
\end{center}
\end{figure}

\noindent The tidal interactions between the disc and secondary cause the secondary to be captured on an eccentric orbit: this facilitates study of the repitition of this outburst event over several orbits.   Figure \ref{fig:mdot_1long} shows the evolution of the accretion rate over three periastra passages; it can be seen that the magnitude of the outburst reduces significantly with each passage.  The ability of the secondary to strip mass from the outer disc diminishes with each passage, partially due to the depletion of the outer disc itself, as well as the secondary disc's influence in regulating mass flow.  Without mass stripping, the secondary cannot accrete, and the disc is not obliged to significantly readjust its mass distribution, preventing the primary accretion event.  This would suggest that these outbursts are not easily repeatable.

\subsection{Simulation 2 - Close Periastron}

\begin{figure}
\begin{center}
\includegraphics[scale = 0.5]{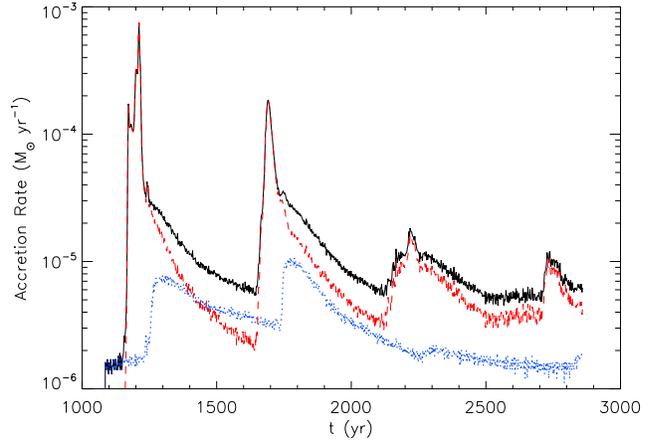} \\
\caption{Accretion rates for the primary and secondary in Simulation 2.  The black line denotes total mass accretion, the blue line indicates the primary accretion, the red line indicates secondary accretion.  \label{fig:mdot_2}}
\end{center}
\end{figure}

\noindent Can repeatability be attained by reducing the secondary periastron? It is not unreasonable to assume that a closer approach allows the secondary to accrete from a more plentiful mass supply, potentially improving its ability to create repeatable outbursts.  The secondary reaches periastron at t \(\sim\) 1155 yr, resulting in the first spike in Figure \ref{fig:mdot_2}.  Feature (ii) is less prominent, as the secondary passes through weaker spiral structure in the inner regions.  Again, the peak accretion rate occurs when the secondary disc begins its infall (feature (iii)), giving a similar order of magnitude increase in accretion rate as Simulation 1. 

The second encounter is of similar magnitude, with an accretion peak at t \(\sim\) 1691 yr.  Note that the accretion rate peak is smoother than the first: the spiral structure in the primary disc has been almost completely erased, so the secondary sees a smooth distribution of mass along its trajectory.  Also, the secondary disc can maintain its shape during the encounter, reducing the infall onto the secondary.  These facts combined eliminate the possibility of seeing features (ii) and (iii) in the second peak: all that remains is the smooth component (feature (i)).  However, this magnitude of outburst cannot be maintained through subsequent orbits, decaying in similar fashion to Simulation 1.  Repeatability therefore seems to be limited for outbursts of this type.

\subsection{Simulation 3 - A More Massive Disc}

\begin{figure}
\begin{center}
\includegraphics[scale = 0.5]{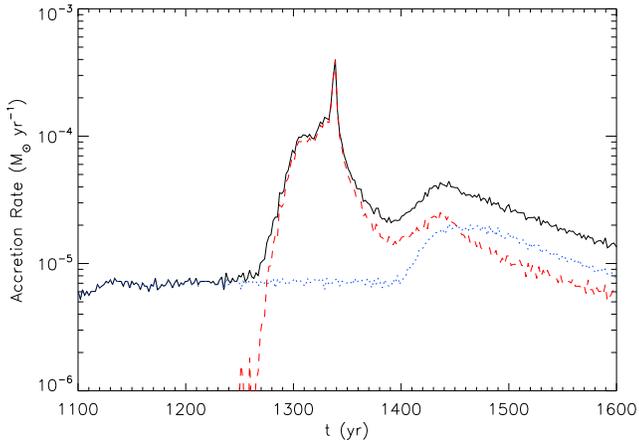} \\
\caption{Accretion rates for the primary and secondary in Simulation 3.  The black line denotes total mass accretion, the blue line indicates the primary accretion, the red line indicates secondary accretion.  \label{fig:mdot_3}}
\end{center}
\end{figure}

\noindent As the important factor is enhanced accretion, will adding more mass to the disc result in a larger accretion event? The results of Simulation 3 (where the disc is double the mass of Simulation 1) indicate some important differences (Figure \ref{fig:mdot_3}).  The disc's lack of high-m spiral modes compared to the lower mass disc (see Figure \ref{fig:disc}) prevents feature (ii) from being presented here.  Periastron occurs at t \(\sim\) 1284 yr, where no significant accretion peak can be seen.  The peak accretion rate is again seen when secondary disc infall occurs at t \(\sim\) 1338 yr.  Overall, the features of this outburst event are similar in magnitude and duration to those seen in the simulations with a less massive disc, in particular the peak accretion rate of the secondary and its decay timescale.  This would suggest that the behaviour of the secondary during the outburst is relatively insensitive to disc mass (possibly because the secondary is already operating at peak accretion efficiency at lower disc masses).  With that said, the pre-outburst and post-outburst accretion rates of the primary are slightly higher, and its decay timescale is slightly shorter.  Depending on the orientation of this system to the observer, this will have important implications for observation (see Discussion).

\subsection{Simulation 4 - A Hyperbolic Encounter}

\begin{figure}
\begin{center}
\includegraphics[scale = 0.5]{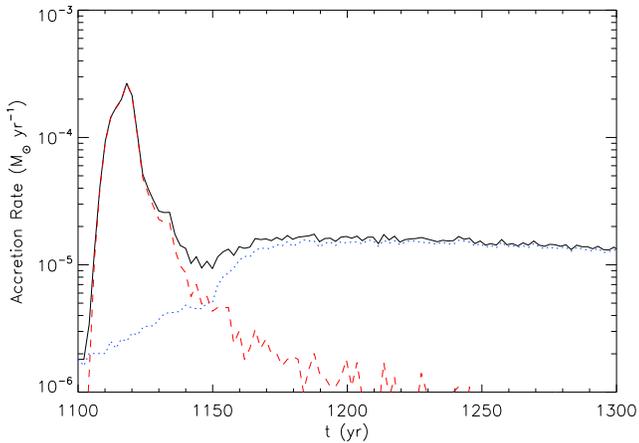} \\
\caption{Accretion rates for the primary and secondary in Simulation 4.  The black line denotes total mass accretion, the blue line indicates the primary accretion, the red line indicates secondary accretion.  \label{fig:mdot_4}}
\end{center}
\end{figure}

\noindent The previous sections of this paper have indicated that accretion efficiency of the secondary is linked to the velocity of the secondary relative to the disc.  If the secondary moves through the disc too quickly, it may be expected that the accretion event will be reduced in magnitude in comparison to the other simulations shown.  Simulation 4 was run to discover the effects of increased orbital velocity, by specifying a hyperbolic encounter ($e=7$).  Figure \ref{fig:mdot_4} indicates that the accretion efficiency is indeed reduced.  The peak accretion rate is lower, and the duration of the accretion event is also reduced, lasting around 50 years.  There is no evidence of feature (ii): the high velocity of the secondary's motion (and the limited time resolution of the data) prevent the detection of these peaks.  There is no peak associated with the secondary's periastron (t \(\sim\) 1110 yr). 

The reduced peak is again associated with the infall of matter onto the secondary after the encounter (feature (iii)), but no disc is formed.  The high-velocity encounter essentially destroys the disc, throwing significant amounts of material to large distances.  It is this halo of matter which the secondary accretes from, but not efficiently or for any length of time, as the velocity dispersion of the material is quite large.

The primary's accretion behaviour remains similar to the other simulations, despite the dispersive action of the secondary.  Indeed, the act of removing disc material may help in the detection of such an event (see Discussion).

\section{Discussion}

\subsection{The Potential for Observation}

\subsubsection{Frequency of Occurence}

\noindent For outbursts from stellar encounters to be observed and correctly classified, their occurence must be sufficiently frequent in an observer's field of view.  To estimate the frequency of stellar encounters in a star cluster, the collision rate calculations of \citet{Clarke_Pringle_binary} are employed.  They calculate the collision rate \(\Gamma_{hit}\) of a star-disc system with discless companions in a star cluster with stellar number density \(n_0\), and a Gaussian velocity distribution, characterised in one dimension by the velocity dispersion \(V_{*}\):

\begin{equation} \Gamma_{hit} = \Gamma_0\left(1+\frac{V^2_{*}R_{disc}}{GM_*} \right) \end{equation}

\noindent where

\begin{equation} \Gamma_0 = \frac{4 \sqrt{\pi}n_0GM_*R_{disc}}{V_{*}} \end{equation}

\noindent The discs used in these simulations has \(R_{disc} = \) 40 au, and \(M_* = 0.5 M_{\odot}\).  To mimic the core of an open cluster, the free parameters are selected to be \(n_{0} = 100\,pc^{-3}\), \(V_{*} = 1\, km\,s^{-1}\) \citep{Binney_Tremaine_87,Clarke_Pringle_binary}.  This yields \(\Gamma_0 = 3.01\times 10^{-4} \,Myr^{-1}\), and \(\Gamma_{hit} = 3.28 \times 10^{-4} \, Myr^{-1}\).  Assuming that the average protostellar disc has a lifetime of \(\sim\) 1 Myr, \(\Gamma_{hit}\) is the probability that one disc will undergo an encounter in its lifetime.  Therefore, out of roughly 3000 stars with discs, one will experience an encounter in its lifetime.  This would imply that in an average open cluster, there will be at least one encounter-driven accretion event per Myr.  Assuming that these encounters are randomly distributed in inclination, it should be expected that only around 6\% of these encounters will be coplanar\footnote{For an encounter to be coplanar, the inclination must be lower than the disc's opening angle, defined by its aspect ratio $\frac{H}{R} = \frac{c_s}{\Omega R}$.  This can be justified by rewriting equation \ref{eq:rbound} for an inclined secondary}.  Finally, the duration of these encounters must be taken into account: the primary's accretion rate remains enhanced for time intervals of $\Delta t = 500$ years, which improves the chances of detection somewhat.  If the probability of detecting an event is defined as

\begin{equation} P_{obs} = N_{events}\Delta t = \Gamma_{hit}N_{stars}f_{inc}\Delta t \end{equation}

\noindent With $f_{inc} = 0.06$, and $\Delta t = 500$ yr, then for $N_{stars} = 3000$, this yields $P_{obs} \sim 10^{-4}$.  This shows that encounters of this type should not be frequently observed. This is of course an oversimplification: it does not account for a distribution of disc radii or masses, nor the subclustering that exists in bound systems.  It also does not reflect the young ages of most FU Ori systems (in general less than half the disc's lifetime).  However, it provides a sufficient order-of-magnitude estimation to illustrate the rarity of these events. 

\subsubsection{The Problems of Obscuration}

The accretion luminosities shown in the previous sections are intrinsic luminosities: they do not account for the effects of optical depth.  Consider the case where the disc is face-on to an observer: the primary star resides in a gap at the centre of the disc, and so the optical depth to the observer is relatively low.  The secondary, however, penetrates the disc, accumulating matter and heating the surroundings.  The Rosseland mean optical depth of the secondary at periastron reaches \(\tau \sim 300\).  In this optically thick regime, the secondary is strongly obscured: any detectable emission will be reprocessed and emitted by the disc at longer wavelengths.

The effect of obscuration increases with inclination to the observer.  The optical depth of the primary (and secondary) to the observer can increase to \(\tau\sim 10^7\) at edge-on, completely shielding the accretion event from observers.  This would suggest that encounters have to occur within a restricted range of inclinations to the observer in order to be directly observable.

\subsubsection{Indirect Observational Signatures}

If the secondary's accretion event is mostly screened by the disc, then what observational signatures can be identified? The disc SED for Simulation 1 was calculated by assuming that the disc emits a blackbody spectrum at each annulus, and integrating the contributions:

\begin{equation} F_{\lambda} = \int 2\pi r B_{\lambda}(T_{eff}(r)) dr\end{equation}

The effective temperature of the spectrum is defined using midplane variables:

\begin{equation} T_{eff}(r) = \frac{T^4_{mid}(r)}{\tau_{mid}(r) + \tau_{mid}^{-1}(r)} \end{equation}

The effect of the encounter is to steepen the surface density profile, which in turn steepens the optical depth profile.  This allows the cooler outer regions to radiate more efficiently, while the inner regions experience stronger screening.  The result is a boost in the disc's flux at longer wavelengths, increasing the flux at \(850 \mu m\) (for example) by a factor of 10.  Add to this the obscuration of the secondary (especially its shorter wavelengths), and it appears that the outburst event is best searched for in the far-IR to sub-mm regimes.

However, the effects of stellar irradiation are not included in the simulation.  Can a substantial increase in reprocessed emission and scattered light be expected from the encounter?  Stellar irradiation (and reprocessed emission) can be estimated as a function of temperature, assuming radiative equilibrium \citep{Ida_and_Lin_1}:

\begin{equation} T(r) \approx 280\left(\frac{r}{1\,AU} \right)^{-\frac{1}{2}} \left(\frac{L_*}{L_{\odot}}\right)^{\frac{1}{4}} \end{equation}

In the initial stage, the primary is the sole contributor: the primary's initial accretion luminosity (plus an intrinsic luminosity of a similar mass main-sequence star) gives

\begin{equation} T(r) \approx 300\left(\frac{r}{1\,AU} \right)^{-\frac{1}{2}} \end{equation}

By comparison, the disc reaches temperatures of over 1000 K in the inner 3 AU initially (from dynamical heating sustained by marginal instability), so it is expected that reprocessed emission as a result of accretion should not be significant initially.  During the encounter, the secondary's accretion luminosity increases to values of \(\sim 40 L_{\odot}\).  This implies that stellar irradiation should be responsible for temperatures  of around 700 K at 1 AU from the secondary.  Compressive heating increases the temperature in this region to a similar value: this suggests that reprocessed starlight will play a significant role, and should also be considered in terms of observations.

\section{Conclusions}\label{sec:conclusions}

\noindent This paper has investigated the possibility of a stellar encounter (where one participant has a protostellar disc) being the progenitor of an outburst phenomenon.  The outbursts described here have several key features in common, originating from two distinct accretion events (for the primary and secondary respectively), independent of the secondary's orbital parameters.  The secondary's accretion rate grows and fluctuates as it traverses the disc's spiral structure: the peak occurs when the secondary disc is being formed from infalling stripped primary disc material, and the accretion rate then decays over several hundred years.  The primary accretion rate (perhaps the easiest to observe) increases slowly with a rise timescale of tens of years as the primary disc readjusts to mass stripping, and decays on a longer timescale than the secondary. 

It has been established that these stellar encounters can enhance accretion rates to levels corresponding to FU Ori and EX Lupi outbursts, and that they have  observational signatures, that although low in probability to detect, are nonetheless possible in principle to see.  But, it is important to emphasise that these encounters cannot be responsible for \emph{all} FU Ori phenomena (or EX Lupi for that matter), for the following reasons:

\begin{enumerate}
\item This type of stellar encounter is too infrequent to explain the catalogue of outbursts currently observed,
\item Such encounters cannot maintain the rapid periodicity or repeatability required of some outbursts without significant decay of the outburst strength,
\item This origin would predict the detection of a companion (perhaps in the infrared or submillimetre) for every outburst host.  This is not the case; out of at least twenty FUors, only seven have a confirmed companion \citep{Pfalzner_08}.
\end{enumerate}

\noindent Despite this, outbursts from stellar encounters can mimic FU Ori well, with the correct general behavioural trends.  They may also potentially have the same triggering mechanism - mass pile up leading to MRI activation \citep{Armitage_et_al_01,Zhu_et_al_09}, although higher resolution simulations (which resolve the inner region and reduce the artificial viscosity) are required to confirm this. Taking the fraction of FUors with a companion as a guide, it is estimated that at most 30\% of outbursts detected could be due to an encounter or binary (with the actual figure presumably much smaller). The outbursts identified here represent a subtly different type of object, that although not currently detected, may be detected in future large-scale surveys of star-forming regions at far-infrared and sub-mm wavelength.

\section*{Acknowledgements}

\noindent All simulations presented in this work were carried out using high performance computing funded by the Scottish Universities Physics Alliance (SUPA).  Surface density plots were made using \begin{small}{SPLASH}\end{small} \citep{SPLASH}.  The authors would like to thank Philip Armitage for useful discussions which helped to refine this work.

\bibliographystyle{mn2e} 
\bibliography{enc_outburst}

\begin{thebibliography}{}

\bibitem[\protect\citeauthoryear{{Armitage}, {Livio} \& {Pringle}}{{Armitage}
  et~al.}{2001}]{Armitage_et_al_01}
{Armitage} P.~J.,  {Livio} M.,    {Pringle} J.~E.,  2001, \mnras, 324, 705

\bibitem[\protect\citeauthoryear{{Bate}}{{Bate}}{2009}]{Bate_09}
{Bate} M.~R.,  2009, \mnras, 392, 1363

\bibitem[\protect\citeauthoryear{{Bate}, {Bonnell} \& {Bromm}}{{Bate}
  et~al.}{2003}]{Bate_cluster_03}
{Bate} M.~R.,  {Bonnell} I.~A.,    {Bromm} V.,  2003, \mnras, 339, 577

\bibitem[\protect\citeauthoryear{{Bate}, {Bonnell} \& {Price}}{{Bate}
  et~al.}{1995}]{Bate_code}
{Bate} M.~R.,  {Bonnell} I.~A.,    {Price} N.~M.,  1995, \mnras, 277, 362

\bibitem[\protect\citeauthoryear{{Bate} \& {Burkert}}{{Bate} \&
  {Burkert}}{1997}]{Burkert_Jeans}
{Bate} M.~R.,  {Burkert} A.,  1997, \mnras, 288, 1060

\bibitem[\protect\citeauthoryear{{Bell} \& {Lin}}{{Bell} \&
  {Lin}}{1994}]{Bell_and_Lin}
{Bell} K.~R.,  {Lin} D.~N.~C.,  1994, \apj, 427, 987

\bibitem[\protect\citeauthoryear{{Binney} \& {Tremaine}}{{Binney} \&
  {Tremaine}}{1987}]{Binney_Tremaine_87}
{Binney} J.,  {Tremaine} S.,  1987, {Galactic Dynamics}.
{Princeton University Press}

\bibitem[\protect\citeauthoryear{{Boley} \& {Durisen}}{{Boley} \&
  {Durisen}}{2008}]{Boley_and_Durisen_09}
{Boley} A.~C.,  {Durisen} R.~H.,  2008, \apj, 685, 1193

\bibitem[\protect\citeauthoryear{{Boley}, {Hartquist}, {Durisen} \&
  {Michael}}{{Boley} et~al.}{2007}]{Boley_hydrogen}
{Boley} A.~C.,  {Hartquist} T.~W.,  {Durisen} R.~H.,    {Michael} S.,  2007,
  \apjl, 656, L89

\bibitem[\protect\citeauthoryear{{Bonnell} \& {Bastien}}{{Bonnell} \&
  {Bastien}}{1992}]{Bonnell_Bastien_92}
{Bonnell} I.,  {Bastien} P.,  1992, \apjl, 401, L31

\bibitem[\protect\citeauthoryear{{Clarke}}{{Clarke}}{2009}]{Clarke_09}
{Clarke} C.~J.,  2009, \mnras, 396, 1066

\bibitem[\protect\citeauthoryear{{Clarke} \& {Pringle}}{{Clarke} \&
  {Pringle}}{1991}]{Clarke_Pringle_binary}
{Clarke} C.~J.,  {Pringle} J.~E.,  1991, \mnras, 249, 584

\bibitem[\protect\citeauthoryear{{Clarke} \& {Syer}}{{Clarke} \&
  {Syer}}{1996}]{Clarke_Syer_96}
{Clarke} C.~J.,  {Syer} D.,  1996, \mnras, 278, L23

\bibitem[\protect\citeauthoryear{{Forgan} \& {Rice}}{{Forgan} \&
  {Rice}}{2009}]{encounters}
{Forgan} D.,  {Rice} K.,  2009, \mnras, in press

\bibitem[\protect\citeauthoryear{{Forgan}, {Rice}, {Stamatellos} \&
  {Whitworth}}{{Forgan} et~al.}{2009}]{intro_hybrid}
{Forgan} D.,  {Rice} K.,  {Stamatellos} D.,    {Whitworth} A.,  2009, \mnras,
  394, 882

\bibitem[\protect\citeauthoryear{{Furlan}, {McClure}, {Calvet}, {Hartmann},
  {D'Alessio}, {Forrest}, {Watson}, {Uchida}, {Sargent}, {Green} \&
  {Herter}}{{Furlan} et~al.}{2008}]{Furlan_et_al_08}
{Furlan} E.,  {McClure} M.,  {Calvet} N.,  {Hartmann} L.,  {D'Alessio} P.,
  {Forrest} W.~J.,  {Watson} D.~M.,  {Uchida} K.~I.,  {Sargent} B.,  {Green}
  J.~D.,    {Herter} T.~L.,  2008, \apjs, 176, 184

\bibitem[\protect\citeauthoryear{{Gingold} \& {Monaghan}}{{Gingold} \&
  {Monaghan}}{1977}]{Gingold_Monaghan}
{Gingold} R.~A.,  {Monaghan} J.~J.,  1977, \mnras, 181, 375

\bibitem[\protect\citeauthoryear{{Hartmann} \& {Kenyon}}{{Hartmann} \&
  {Kenyon}}{1996}]{Hartmann_Kenyon_96}
{Hartmann} L.,  {Kenyon} S.~J.,  1996, \araa, 34, 207

\bibitem[\protect\citeauthoryear{{Herbig}}{{Herbig}}{1977}]{Herbig_77}
{Herbig} G.~H.,  1977, \apj, 217, 693

\bibitem[\protect\citeauthoryear{{Herbig}}{{Herbig}}{2007}]{Herbig_07}
{Herbig} G.~H.,  2007, \aj, 133, 2679

\bibitem[\protect\citeauthoryear{{Ida} \& {Lin}}{{Ida} \&
  {Lin}}{2004}]{Ida_and_Lin_1}
{Ida} S.,  {Lin} D.~N.~C.,  2004, \apj, 604, 388

\bibitem[\protect\citeauthoryear{{Kenyon}, {Calvet} \& {Hartmann}}{{Kenyon}
  et~al.}{1993}]{Kenyon_et_al_93}
{Kenyon} S.~J.,  {Calvet} N.,    {Hartmann} L.,  1993, \apj, 414, 676

\bibitem[\protect\citeauthoryear{{Kenyon}, {Hartmann} \& {Hewett}}{{Kenyon}
  et~al.}{1988}]{Kenyon_et_al_88}
{Kenyon} S.~J.,  {Hartmann} L.,    {Hewett} R.,  1988, \apj, 325, 231

\bibitem[\protect\citeauthoryear{{Kenyon}, {Hartmann}, {Strom} \&
  {Strom}}{{Kenyon} et~al.}{1990}]{Kenyon_et_al_90}
{Kenyon} S.~J.,  {Hartmann} L.~W.,  {Strom} K.~M.,    {Strom} S.~E.,  1990,
  \aj, 99, 869

\bibitem[\protect\citeauthoryear{{Lin}, {Faulkner} \& {Papaloizou}}{{Lin}
  et~al.}{1985}]{Lin_Papa_85}
{Lin} D.~N.~C.,  {Faulkner} J.,    {Papaloizou} J.,  1985, \mnras, 212, 105

\bibitem[\protect\citeauthoryear{{Lodato} \& {Clarke}}{{Lodato} \&
  {Clarke}}{2004}]{Lodato_Clarke_04}
{Lodato} G.,  {Clarke} C.~J.,  2004, \mnras, 353, 841

\bibitem[\protect\citeauthoryear{{Lodato}, {Meru}, {Clarke} \& {Rice}}{{Lodato}
  et~al.}{2007}]{Lodato_encounters}
{Lodato} G.,  {Meru} F.,  {Clarke} C.~J.,    {Rice} W.~K.~M.,  2007, \mnras,
  374, 590

\bibitem[\protect\citeauthoryear{{Lodato} \& {Rice}}{{Lodato} \&
  {Rice}}{2004}]{Lodato_and_Rice_04}
{Lodato} G.,  {Rice} W.~K.~M.,  2004, \mnras, 351, 630

\bibitem[\protect\citeauthoryear{{Lucy}}{{Lucy}}{1977}]{Lucy}
{Lucy} L.~B.,  1977, \aj, 82, 1013

\bibitem[\protect\citeauthoryear{{Mayer}, {Lufkin}, {Quinn} \&
  {Wadsley}}{{Mayer} et~al.}{2007}]{Mayer_et_al_07}
{Mayer} L.,  {Lufkin} G.,  {Quinn} T.,    {Wadsley} J.,  2007, \apjl, 661, L77

\bibitem[\protect\citeauthoryear{{Monaghan}}{{Monaghan}}{1992}]{Monaghan_92}
{Monaghan} J.~J.,  1992, \araa, 30, 543

\bibitem[\protect\citeauthoryear{{Monaghan}}{{Monaghan}}{2005}]{Monaghan_05}
{Monaghan} J.~J.,  2005, \repprog, 68, 1703

\bibitem[\protect\citeauthoryear{{Pfalzner}}{{Pfalzner}}{2008}]{Pfalzner_08}
{Pfalzner} S.,  2008, \aap, 492, 735

\bibitem[\protect\citeauthoryear{{Price}}{{Price}}{2007}]{SPLASH}
{Price} D.~J.,  2007, Publications of the Astronomical Society of Australia,
  24, 159

\bibitem[\protect\citeauthoryear{{Rice} \& {Armitage}}{{Rice} \&
  {Armitage}}{2009}]{Rice_and_Armitage_09}
{Rice} W.~K.~M.,  {Armitage} P.~J.,  2009, \mnras, pp 709--+

\bibitem[\protect\citeauthoryear{{Rodr{\'{\i}}guez}, {Loinard}, {D'Alessio},
  {Wilner} \& {Ho}}{{Rodr{\'{\i}}guez} et~al.}{2005}]{Rodriguez_et_al_05}
{Rodr{\'{\i}}guez} L.~F.,  {Loinard} L.,  {D'Alessio} P.,  {Wilner} D.~J.,
  {Ho} P.~T.~P.,  2005, \apjl, 621, L133

\bibitem[\protect\citeauthoryear{{Shakura} \& {Sunyaev}}{{Shakura} \&
  {Sunyaev}}{1973}]{Shakura_Sunyaev_73}
{Shakura} N.~I.,  {Sunyaev} R.~A.,  1973, \aap, 24, 337

\bibitem[\protect\citeauthoryear{{Shu}, {Adams} \& {Lizano}}{{Shu}
  et~al.}{1987}]{Shu_et_al_87}
{Shu} F.~H.,  {Adams} F.~C.,    {Lizano} S.,  1987, \araa, 25, 23

\bibitem[\protect\citeauthoryear{{Stamatellos}, {Hubber} \&
  {Whitworth}}{{Stamatellos} et~al.}{2007}]{Stam_frag}
{Stamatellos} D.,  {Hubber} D.~A.,    {Whitworth} A.~P.,  2007, \mnras, 382,
  L30

\bibitem[\protect\citeauthoryear{{Stamatellos}, {Whitworth}, {Bisbas} \&
  {Goodwin}}{{Stamatellos} et~al.}{2007}]{Stam_2007}
{Stamatellos} D.,  {Whitworth} A.~P.,  {Bisbas} T.,    {Goodwin} S.,  2007,
  \aap, 475, 37

\bibitem[\protect\citeauthoryear{{Vorobyov} \& {Basu}}{{Vorobyov} \&
  {Basu}}{2005}]{Vorobyov_Basu_05}
{Vorobyov} E.~I.,  {Basu} S.,  2005, \apjl, 633, L137

\bibitem[\protect\citeauthoryear{{Vorobyov} \& {Basu}}{{Vorobyov} \&
  {Basu}}{2006}]{Vorobyov_Basu_06}
{Vorobyov} E.~I.,  {Basu} S.,  2006, \apj, 650, 956

\bibitem[\protect\citeauthoryear{{Vorobyov} \& {Basu}}{{Vorobyov} \&
  {Basu}}{2008}]{Vorobyov_Basu_08}
{Vorobyov} E.~I.,  {Basu} S.,  2008, \apjl, 676, L139

\bibitem[\protect\citeauthoryear{{Whitehouse} \& {Bate}}{{Whitehouse} \&
  {Bate}}{2004}]{WB_1}
{Whitehouse} S.~C.,  {Bate} M.~R.,  2004, \mnras, 353, 1078

\bibitem[\protect\citeauthoryear{{Zhu}, {Hartmann} \& {Gammie}}{{Zhu}
  et~al.}{2009}]{Zhu_et_al_09}
{Zhu} Z.,  {Hartmann} L.,    {Gammie} C.,  2009, \apj, 694, 1045

\end{thebibliography}

\label{lastpage}

\end{document}